\documentclass[11pt]{article}
\date{today}
\usepackage[dvips]{graphicx}
\usepackage{latexsym}
\usepackage{euscript}

\newcommand{\Tr}{\mbox{Tr}\;}

\newcommand{\<}{\langle}
\renewcommand{\>}{\rangle}
\newcommand{\be}{\begin{equation}}
\newcommand{\ee}{\end{equation}}
\newcommand{\ba}{\begin{eqnarray}}
\newcommand{\ea}{\end{eqnarray}}

\newcommand{\beq}{\begin{equation}}
\newcommand{\eeq}{\end{equation}}
\newcommand{\beqn}{\begin{eqnarray}}
\newcommand{\eeqn}{\end{eqnarray}}
\begin{document}
\pagestyle{empty}
\vspace{-0.6in}
\begin{flushright}
CPT-2004/P.006\\
ROME1-1370/2004\\
ROM2F/2004/02\\
\end{flushright}
\vskip 1.5in

\centerline{\Large {\bf{Topological susceptibility in full QCD}}}
\medskip
\centerline{\Large {\bf{with Ginsparg-Wilson fermions}}}
\vskip 0.6cm
\centerline{{L. Giusti}$^{a}$, {G.C. Rossi}$^{b}$, {M. Testa}$^{c}$}
\vskip 0.5cm
\centerline{$^a$ Centre de Physique Th\'eorique, CNRS Luminy, Case 907,
F-13288 Marseille, France} \vskip 0.2cm \centerline{$^b$
Dipartimento di Fisica, Universit\`a di Roma ``{\it Tor Vergata}''}
\centerline{INFN, Sezione di Roma 2} \centerline{Via della Ricerca
Scientifica, I-00133 Roma, Italy} \vskip 0.2cm \centerline{$^c$
Dipartimento di Fisica, Universit\`a di Roma ``{\it La Sapienza}"}
\centerline{INFN, Sezione di Roma ``{\it La Sapienza}''}
\centerline{P.le A. Moro 2, I-00185 Roma, Italy}

\vskip 1.0in
\begin{abstract}
We show that, if the formula for the topological charge density
operator suggested by fermions obeying the Ginsparg--Wilson
relation is employed, it is possible to give a precise and
unambiguous definition of the topological susceptibility in full QCD, 
$\chi^{\rm{full}}_{t{\rm{L}}}$, for finite quark masses on the lattice. The 
lattice expression of $\chi^{\rm{full}}_{t{\rm{L}}}$ looks like the formal 
continuum one, in the sense that no power divergent subtractions are needed
for its proper definition. As a consequence, the small mass behaviour 
of $\chi^{\rm{full}}_{t{\rm{L}}}$ leads directly to a multiplicative 
renormalizable definition of the chiral condensate that does not require any
power divergent subtraction.
\end{abstract}
\begin{flushleft}
\end{flushleft}
\vfill
\pagestyle{empty}\clearpage
\setcounter{page}{1}
\pagestyle{plain}
\newpage
\pagestyle{plain} \setcounter{page}{1}

\newpage

\section{Introduction}
\label{sec:INTRO}
In this paper we discuss the definition and the properties of the 
topological susceptibility in full QCD with massive quarks 
extending the results of ref.~\cite{GRTV}. Using
arguments based on anomalous flavour singlet Ward-Takahashi
identities (WTI's), we prove that, if the formula
of the topological charge density, $Q$, suggested by Ginsparg--Wilson
(GW) fermions~\cite{Ginsparg:1982bj,NaraNeub,Neuberger} is
employed, the full QCD topological susceptibility 
\be\label{eq:CHIT}
\chi_{tL}^{\rm{full}}=\int d^4x\,\<Q(x)Q(0)\> 
\ee 
does not need any power divergent subtraction 
at finite non-vanishing values of the quark 
(pion) mass. Furthermore it vanishes linearly in the quark mass with a
coefficient which turns out to be (proportional to) the chiral
condensate, as in the formal continuum limit.

The interest of these results lies in the fact that one can exploit
the absence of power divergent mixings in the continuum-looking lattice
formula~(\ref{eq:CHIT}) to extract the value of the chiral condensate
with no need of performing any dangerous power divergent subtraction.

\section{Generalities on Ginsparg-Wilson fermions}
\label{sec:GEN}

Regularizing the fermionic part of the QCD action using GW fermions offers
the great advantage that
global chiral transformations can be defined, which are an exact symmetry
of the massless theory, as in the formal continuum theory. This is a
consequence of the relation~\cite{Ginsparg:1982bj}
\beq\label{eq:GW}
\gamma_5 D + D \gamma_5 = a  D \gamma_5 D\, ,
\eeq
where $D$ is the Dirac operator and $a$ is the lattice spacing. 
Eq.~(\ref{eq:GW}) implies the invariance of the massless fermion
action under the transformations~\cite{Luscher:1998pq}
\beq\label{eq:luscher_new}
\delta_A^f \psi = \lambda^f \widehat\gamma_5 \psi\, ,
\qquad \delta_A^f \bar\psi =
\bar\psi \gamma_5 \lambda^f\, ,\quad f=0,1,\ldots,N_f^2-1  \, ,
\eeq
where $\lambda^0=1\!\!1$ and the $\lambda^f$'s, $f\neq 0$, are 
flavour matrices~\footnote{We use the normalization
${\rm{tr}}(\lambda^f\lambda^{g})=\delta^{fg}/2$,
$[\lambda^f,\lambda^{g}]=i f^{fgh}\lambda^h$, so that
$\{\lambda^f,\lambda^{g}\}=d^{fgh}\lambda^h+\delta^{fg}1\!\!1/N_f$,
$f,g,h=1,\ldots,N_f^2-1$.}. In the first of the equations above we have 
introduced the definition
\beq\label{eq:GAMMA5}
\widehat \gamma_5 = \gamma_5\Big(1- a D\Big)
\eeq
with the properties
\beq
\widehat\gamma_5^\dagger=\widehat\gamma_5\, ,\qquad\widehat\gamma_5^2 = 1\, .
\eeq
Eqs.~(\ref{eq:luscher_new}) may be interpreted as the lattice form of the
continuum chiral transformations. The Neuberger operator~\cite{Neuberger} 
satisfies the GW relation, has the correct continuum limit and is local, 
though not ultra-local~\cite{Hernandez:1998et}.
Another solution of the GW condition~(\ref{eq:GW}) is provided by the 
fixed-point fermionic action of refs.~\cite{PerfectA,Hasenfratz:1997ft}.

In a GW regularization the lattice QCD fermion action with $N_f$ massive
flavour can be written in the form~\footnote{For short we use continuum
looking notations with $\int d^4x$ replacing $a^4\sum_x$.}
\beqn\label{eq:action}
S_F = \int d^4x \sum_{r,s=1}^{N_f}\bar{\psi}^r(x)\Big[\left(D\delta_{rs}+ 
P_- M_{rs}^\dagger
\widehat P_- + P_+ M_{rs} \widehat P_+\right)\psi^s \Big](x)\label{eq:SACT}
\eeqn
where
\beq
\widehat P_{\pm} = \frac{1}{2}(1\pm \widehat \gamma_5)\, ,\quad
P_{\pm} = \frac{1}{2}(1\pm \gamma_5)\, ,
\eeq
$\psi$ ($\bar\psi$) is an $N_f$-dimensional column (row) vector in flavour
space and $M=\mbox{diag}\,(m_1,\dots , m_{N_f})$. $S_F$ is invariant
under the $U_{L}(N_f)\times U_{R}(N_f)$ global transformations
\beqn\label{eq:chitransf}
\psi_L \rightarrow U_L \psi_L\;  & \qquad &
\bar \psi_L \rightarrow \bar \psi_L U_L^{\dagger}\nonumber\\
\psi_R \rightarrow U_R \psi_R\;  & \qquad &
\bar \psi_R \rightarrow \bar \psi_R U_R^{\dagger}\, ,
\eeqn
with $U_{L,R}\in U(N_f)_{L,R}$ and
\beqn\label{eq:transf}
\psi_L = \widehat P_{-}\psi\;  & \qquad &
\bar \psi_L = \bar \psi P_{+} \nonumber\\
\psi_R = \widehat P_{+}\psi\;  & \qquad & \bar \psi_R = \bar \psi
P_{-} \, , \eeqn 
if at the same time $M\rightarrow U_L M U_R^\dagger$. In the 
following we shall restrict ourselves to the flavour vector symmetric case 
$m_r=m$, with $r=1,\ldots,N_f$. As a consequence of the exact chiral 
invariance of the massless GW regularization, no additive quark mass 
renormalization is required. The action is O$(a)$-improved, since no chiral 
invariant operators of dimension $d=5$ can be constructed. In this work we 
will consider the bilinear scalar and pseudo-scalar quark operators 
($f=0,1,\dots,N_f^2-1$)
\beqn
&&S^{f}(x) =\bar \psi(x)\lambda^f 
\Bigl[\left(1-\frac{a}{2}D\right)\psi\Bigr](x)\, ,\label{eq:SS} \\
&&P^{f}(x) =\bar \psi(x)\lambda^f\gamma_5 
\Bigl[\left(1-\frac{a}{2}D\right)\psi\Bigr](x)\, .\label{eq:PP} 
\eeqn  
The ``rotation'' $(1-\frac{a}{2}D)$ of the 
quark field $\psi$ in the above equations leads
to definitions of scalar and  pseudo-scalar quark densities
which have the correct chiral transformation
properties, like in the formal continuum theory, and only need a 
(logarithmically divergent) multiplicative renormalization. Furthermore 
the operators~(\ref{eq:SS}) and~(\ref{eq:PP}) are automatically 
O$(a)$-improved. 

In a GW regularization the gauge anomaly is recovered {\it \`a la}
Fujikawa~\cite{Fujikawa:1979ay,Luscher:1998pq}. The fermion
integration measure is not invariant under $U_A(1)$
transformations (eqs.~(\ref{eq:luscher_new}) with $f=0$), 
and the topological charge density
\beq
\label{eq:TCDO} a^4 Q(x) = -\frac{a}{2}\Tr\Big[\gamma_5
D(x,x)\Big]\, , 
\eeq 
originating from the corresponding Jacobian,
is related to the index of the lattice Dirac operator, $D$, 
by the equation~\cite{Neuberger,Hasenfratz:1998ri,Luscher:1998pq} 
\beq n_+ - n_-={\rm{index}}\,(D)=\int d^4x\, Q(x) \, , \label{eq:INDEX}
\eeq with $n_+$ ($n_-$) the number of zero modes 
with positive (negative) chirality~\footnote{For alternative
lattice definitions of $Q$ see the papers of ref.~\cite{QTOP}.}.
For a recent review on this subject see~\cite{Giusti:2002rx}.

\section{The singlet  Ward-Takahashi identities}
In the chiral limit, the local anomalous flavour-singlet 
WTI's have the form
\beq
\partial_\mu \<{\cal A}^0_\mu (x) \widehat{\cal O}(y)\>=
2N_f \< Q(x) \widehat{\cal O}(y)\>-\<\delta^{0,x}_A \widehat{\cal O}(y)\> \, ,
\label{eq:AWTIL} 
\eeq 
where ${\cal A}^0_\mu (x)$ is the singlet axial current, 
$\widehat{\cal O}$ is any renormalized (multi-)local operator, 
concentrated at points $y\equiv\{y_i,i=1,\ldots,n\}$ 
and $\delta^{0,x}_A \widehat{\cal O}$ is its local singlet 
axial variation. In eq.~(\ref{eq:AWTIL}) we have not shown 
the exponentially suppressed terms coming from the fact that $D$ 
is not ultra-local~\cite{Hernandez:1998et}. They are of no relevance 
for the following arguments, as they vanish after integration. 
Assuming the absence of a $U_A(1)$ massless Goldstone boson, the 
integrated form of the WTI's~(\ref{eq:AWTI}) reads
\beq 
0=2 N_f \int d^4x\,\<Q(x)\widehat{\cal O}(y)\>-\<\delta^0_A
\widehat{\cal O}(y)\>\, . \label{eq:AWTI} 
\eeq 
Since the second term
in the r.h.s.\ of eq.~(\ref{eq:AWTI}) is finite, it follows that
$\int d^4x\,Q(x)$ is also finite, as it has finite insertions with
any string of renormalized fundamental fields. Therefore $Q(x)$
can only mix with operators of dimension $\leq 4$ and vanishing
integral, hence only with $\partial_\mu {\cal A}^0_\mu(x)$. No
power-divergent subtractions with lower dimensional operators
(such as the pseudo-scalar quark density) are required~\cite{GRTV}. 
This is a very distinctive feature of GW
fermions with respect to standard Wilson fermions which is
directly related to the absence of an additive mass 
renormalization~\footnote{A discussion of the singlet WTI's 
for Wilson fermions can be found in ref.~\cite{GS}.}.
One can define finite operators $\widehat Q$ and
$\widehat {\cal A}^0_\mu$ by the equations~\cite{BAR} 
\ba 
\widehat{Q}(x)& = & Q(x) - \frac{Z}{2N_f}\partial_\mu {\cal A}_\mu^0 (x)\, ,
\label{eq:FINQ1}\\
\widehat{\cal A}_\mu^0 (x) & = &(1-Z) {\cal A}_\mu^0 (x)\, , \label{eq:FINA} 
\ea 
where $Z$ is the mixing coefficient between $Q$ and 
$\partial_\mu {\cal A}^0_\mu$. The renormalized singlet axial WTI's then
become (again up to exponentially small terms) \beq
\partial_\mu\<\widehat {\cal A}^0_\mu (x) \widehat{\cal O}(y)\>=
2N_f\<\widehat Q(x) \widehat{\cal O}(y)\>-\<\delta^{0,x}_A 
\widehat{\cal O}(y)\> \, .\label{eq:WTIREN}
\eeq

Outside of the chiral limit the integrated singlet axial WTI's read
\begin{equation}
0=2 m\int d^4x\,\langle P^0(x) \widehat{\cal O}(y) \rangle +
2 N_f \int d^4x\,\langle Q(x) \widehat{\cal O}(y)\rangle -
\int d^4x\,\langle \delta^{0,x}_A \widehat{\cal O}(y)\rangle \, .
\label{eq:AWI_O}
\end{equation}
The extra term present in eq.~(\ref{eq:FINQ1}) being a total 
divergence does not contribute to the integrated WTI's~(\ref{eq:AWI_O}). 
If we replace $\widehat{\cal O}(y)$ with the local operator $Q(0)$, we
get \beq 0=2m\int d^4x\,\langle P^0(x) Q(0)\rangle +2N_f\int
d^4x\,\langle Q(x) Q(0)\rangle\, , \label{eq:AWI_Q} 
\eeq 
and, similarly, by inserting the multiplicative renormalizable operator 
$P^0(0)$ 
\beq 0=2m\int d^4x\,\langle
P^0(x)P^0(0)\rangle+ 2N_f\int d^4x\,\langle Q(x) P^0(0)\rangle
-2\langle S^0(0) \rangle\, . \label{eq:AWI_P0} 
\eeq 
Putting together eqs.~(\ref{eq:AWI_Q}) and~(\ref{eq:AWI_P0}), we obtain
\beq\label{eq:AWI_P0Q} \chi_{tL}^{\rm{full}}\equiv\int d^4x\,\langle Q(x)
Q(0)\rangle = \frac{(2 m)^2}{(2N_f)^2} \int d^4x\,\langle P^0(x)
P^0(0) \rangle - \frac{4m}{(2N_f)^2} \langle S^0(0)\rangle \; .
\eeq
In the next section we show that the full QCD topological susceptibility,
$\chi_{tL}^{\rm{full}}$, defined above is not affected by power
divergences. 

\subsection{Absence of power divergences in $\chi_{tL}^{\rm{full}}$}
\label{sec:ABS}

The proof of the absence of power divergences in $\chi_t^{\rm{full}}$
is based on the study of the short distance beahviour of the
two terms in the r.h.s.\ of eq.~(\ref{eq:AWI_P0Q}) at small $m$.

The argument goes through a number of steps. First of all we observe that
thanks to the chiral properties of GW fermions, only power divergences
of the type $m^2/a^2$ can possibly be present in eq.~(\ref{eq:AWI_P0Q}).
The second observation is that the $m^2/a^2$ divergences separately
affecting the two terms in the r.h.s.\ actually cancel each other. 
This is the result of the exact compensation between the (quadratically) 
divergent term arising in $m^2\int d^4x\,\<P^0(x)P^0(0)\>$, due
to the short distance behaviour of the integrand, and a similar divergent term
appearing in $\<S^0\>$, when one power of the fermionic mass term 
(brought down from the action) is inserted together with
$S^0$. The compensation follows from the fact that the short distance
(perturbative) behaviour of the two correlators $\<P^0(x)P^0(0)\>$ and
$\<S^0(x)S^0(0)\>$ are (in the massless limit) equal (up to a minus sign
due to the presence of two extra $\gamma_5$ matrices in $\<P^0(x)P^0(0)\>$),
leaving behind a finite, computable contribution.

We now make explicit and precise the line of arguments sketched above.
\begin{enumerate}
\item Each term in eq.~(\ref{eq:AWI_P0Q}) is even under the 
(non-anomalous) spurionic symmetry~\cite{FR}
\begin{equation}
{\cal{R}}_5^{\rm{sp}}\equiv{\cal{R}}_5\times [m\rightarrow -m]\qquad
{\cal{R}}_5 : \left \{\begin{array}{ll}
\psi&\rightarrow\psi'=\widehat \gamma_5 \psi  \\\\
\bar{\psi}&\rightarrow\bar{\psi}'=-\bar{\psi}\gamma_5
\end{array}\right .
\label{SPURM0}\end{equation}
where ${\cal{R}}_5$ is an element of the chiral group. Since only 
the identity operator, which is even under ${\cal{R}}_5^{\rm{sp}}$,
can contribute a divergent term in the two terms in the r.h.s.\ of
eq.~(\ref{eq:AWI_P0Q}), we conclude that only 
$m^2/a^2$ power divergences can be present, as they are even under
$m\rightarrow -m$. In other words chiral invariance forbids
power divergences like $m/a^3$ and $m^3/a$.
\item If we order the terms contributing to eq.~(\ref{eq:AWI_P0Q}) 
in powers of $m$, we get
\begin{eqnarray}
\hspace{-1.cm}4N_f^2\chi_{tL}^{\rm{full}}&=&- 4m\<S^0(0)\>\Big{|}_{m=0}+
4m^2\int d^4x\,
\Big[\<S^0(x)S^0(0)\> + \<P^0(x)P^0(0)\> \Big{]}_{m=0}+\label{eq:PROOF}\\
&+ &\,\Big{[}{\mbox{O}}(m^3/m^4)\,\,
{\mbox{with/without {{S$\chi$SB}}}}\Big{]}\, ,\nonumber
\end{eqnarray}
where odd powers of $m$ can be
present in the expansion only if chiral symmetry is spontaneously
broken. For the purpose of studying the structure of power
divergences, we only need to examine terms O($m$) and O($m^2$).
Higher order terms are at most logarithmically
divergent~\footnote{Whether the logarithmically divergent terms
proportional to $m^4$ might be reabsorbed by renormalizing
$m$ is a question that can be decided by a perturbative
calculation.}. 
\item The first term in the expansion~(\ref{eq:PROOF}) is finite. 
In fact, i) owing to the exact chiral symmetry of the massless GW 
fermionic action, there cannot be any mixing between the identity and 
the operator $S^0$ (with a $a^{-3}$ divergent coefficient), because 
they transform in the opposite way under ${\cal{R}}_5^{\rm{sp}}$; ii) 
the quantity $m\<S^0(0)\>|_{m=0}$ is not logarithmically divergent, 
as a consequence of the non-singlet WTI's. 
\item The sum of the next two terms is finite. To prove this result 
it is convenient to consider the set of WTI's ($f=1,\ldots,N_f^2-1$;
$g,h=0,\ldots,N_f^2-1$)
\begin{eqnarray}
\hspace{-1.cm}&&0=\int d^4z\,\int d^4x\,\partial_\mu
\<{\cal{A}}_\mu^f(z) P^g(x)S^{h}(0)\>=\nonumber\\
\hspace{-1.cm}&&=2m\int d^4z\,\int d^4x\,\<P^f(z)P^g(x)S^{h}(0)\>+\nonumber\\
\hspace{-1.cm}&&-\int d^4x\,\<\delta^f_A P^g(x)S^{h}(0)\>-
\int d^4x\,\<P^g(x)\delta^f_A S^{h}(0)\>\, ,\label{eq:TWTIS}
\end{eqnarray}
where $\delta^f_A$ represents the operation of taking the axial variation
with flavour index $f$.
\item Combining the above WTI's, 
one gets in the chiral limit the soft-pion theorem (no sum over 
repeated indices, $d^{fgh}\neq 0$)
\begin{eqnarray}
\hspace{-2.cm}&&\int d^4x\,\Big{[}
\<S^0(x)S^0(0)\>+\<P^0(x)P^0(0)\>\Big{]}_{m=0}=\nonumber\\
\hspace{-2.cm}&&= F_\pi N_f\Big{[}\int d^4x\,
\<\pi^f|T_E(P^f(x)S^0(0)+S^f(x)P^0(0))|0\>\Big{|}_{m=0}+\nonumber\\
\hspace{-2.cm}&&-\frac{2}{d^{fgh}}\int d^4x\,
\<\pi^f|T_E(P^g(x)S^h(0))|0\>\Big{|}_{m=0}\Big{]}
\, ,\label{eq:FORMULA}\end{eqnarray}
where $T_E$ means Euclidean time-ordering~\footnote{For simplicity we do not 
employ a different notation for the operators in the matrix elements appearing
in the r.h.s.\ of eq.~(\ref{eq:FORMULA}).} and we have used the definition
\beq \<0|\partial_\mu{\cal{A}}_\mu^f|\pi^g\>=\delta^{fg}F_\pi m_\pi^2\, .
\label{eq:FP}\eeq 
From the O.P.E.'s
\begin{equation}
P^f(x)S^0(0)\simeq S^f(x)P^0(0)\simeq
\frac{1}{x^3}P^f(0)\, ,\label{eq:OPE}
\end{equation}
\begin{equation}
P^f(x)S^g(0)\simeq \sum_{h}d^{fgh}\frac{1}{x^3}P^h(0)\, ,\label{eq:OPES}
\end{equation}
one concludes that the integrals in the r.h.s.\ of eq.~(\ref{eq:FORMULA}) 
are indeed finite.
\end{enumerate}

\section{Final considerations}
\label{sec:FINC}

A number of interesting consequences follow from
the formul\ae~(\ref{eq:WTIREN}) and~(\ref{eq:AWI_P0Q}).
\begin{enumerate}
\item In the full theory $m^2_{\eta'}\neq 0$ and there is no
massless particle that can couple to $P^0$. So it is immediately
seen that $\chi_{tL}^{\rm{full}}$ vanishes in the chiral limit 
($m\rightarrow 0$). \item A formula for the (quenched) $\eta'$
mass~\cite{WIT,VEN} can be obtained starting from the Fourier
transform of the WTI~(\ref{eq:WTIREN}) at zero quark mass, if one
chooses $\widehat{O}=\widehat{Q}$. For completeness we recall here
the standard argument which goes as follows.
First we observe that the $U_A(1)$ variation of $\widehat{Q}$ is zero.
Taking the Fourier transform of eq.~(\ref{eq:WTIREN}) with $\widehat O$ 
replaced by $\widehat Q$, one gets (in the chiral limit) \beq i\int
d^4x\,e^{-ipx}p_\mu\<\widehat {\cal A}^0_\mu (x) \widehat{Q}(0)\>= 2N_f\int
d^4x\,e^{-ipx}\<\widehat{Q}(x)\widehat{Q}(0)\>\, . \label{eq:WTIFT} \eeq In
the limit $N_f/N_c\rightarrow 0$, where the $\eta'$ mass vanishes,
the l.h.s.\ of eq.~(\ref{eq:WTIFT}) is dominated at small $p$
by the $\eta'$ pole, leading to the expansion \beq i\int
d^4x\,e^{-ipx}p_\mu\<\widehat {\cal A}^0_\mu (x)
\widehat{Q}(0)\>\Big{|}_{N_f/N_c=0} = \lim_{N_f/N_c\rightarrow 0}
m_{\eta'}^2F_{\eta'}^2\frac{p^2}{p^2+m^2_{\eta'}}+ \mbox{O}(p^2)\,
. \label{eq:WTIEXP} \eeq If, as indicated in the above formula,
the limit $p\rightarrow 0$ is taken after the limit
$N_f/N_c\rightarrow 0$, one ends up with the 
relation
\beq
\frac{m_{\eta'}^2F_{\eta'}^2}{2N_f}\Big{|}_{N_f/N_c=0}=
\lim_{p\rightarrow 0}\lim_{N_f/N_c\rightarrow 0} \int d^4x\,
e^{-ipx}\<\widehat{Q}(x)\widehat{Q}(0)\>\, , \label{eq:WTIETA} 
\eeq where
standard counting arguments~\cite{VENTOP} ensures
that the l.h.s.\ of eq.~(\ref{eq:WTIETA}) has a finite limit as
$N_f/N_c\rightarrow 0$. At this point if it is assumed that, taking
the limit ${N_f/N_c\rightarrow 0}$ in the r.h.s.\ of~(\ref{eq:WTIEXP}) 
is equivalent to drop the fermion
determinant, one arrives at the famous WV formula~\footnote{For a 
discussion of several subtleties on the derivation 
of this formula see ref.~\cite{seiler}.} 
\beq
\frac{m_{\eta'}^2F_{\pi}^2}{2N_f}\Big{|}_{N_f/N_c=0}= \int
d^4x\,\<Q(x)Q(0)\>\Big{|}_{\rm{YM}}\, . \label{eq:WVFOR} 
\eeq
Notice that in the limit $N_f/N_c\rightarrow 0$, the mixing
coefficient, $Z$, in eq.~(\ref{eq:FINQ1}) vanishes and $F_{\eta'}$
becomes equal to $F_\pi$. Thus in eq.~(\ref{eq:WVFOR}) we have
replaced $F_{\eta'}$ with $F_\pi$ and at the same time $\widehat{Q}$
with $Q$, although in this limit the integral of the divergence of
the singlet axial current does not vanish. Recalling the
expression~(\ref{eq:TCDO}) and the index theorem~(\ref{eq:INDEX}),
one can equivalently write for the (quenched) $\eta'$ mass the
suggestive formula \beq
\frac{m_{\eta'}^2F_\pi^2}{2N_f}\Big{|}_{N_f/N_c=0}=
\lim_{V\to\infty}\frac{\<(n_+-n_-)^2\>_{V}}{V}\, .
\label{eq:ETAMASSIND}\eeq where $V$ is the space-time volume of
the lattice. \item Paying due care to flavour matrix
normalization, one can combine the non-singlet WTI (written for a
given fixed flavour, $h$) \beq\label{eq:AWI_Pf} 0=2m\int
d^4x\,\langle P^h(x)P^h(0)\rangle-\frac{1}{N_f}\langle
S^0(0)\rangle \eeq with eq.~(\ref{eq:AWI_P0Q}), obtaining \beq (2
N_f)^2 \int d^4x\,\langle Q(x) Q(0)\rangle = (2 m)^2 \int
d^4x\,\langle P^0(x) P^0(0) \rangle\Big{|}^{\rm{ZV}} \, ,
\label{eq:AWI_ZV} \eeq where the superscript ZV means that only
Zweig-Violating (hairpin) diagrams should be included in the
r.h.s.\ of the equation in carrying out the fermion functional
integral. For GW fermions the relation~(\ref{eq:AWI_ZV}) is an
algebraic identity that can be directly proved, using
eq.~(\ref{eq:GW}), the definition~(\ref{eq:TCDO}) and the explicit
expressions of $P^0$ and $S^0$ (eqs.~(\ref{eq:SS}) and~(\ref{eq:PP})). 
A few observations are in order here.
\begin{itemize}
\item An equation like~(\ref{eq:AWI_ZV}) holds also if the fermion 
determinant is neglected (quenching). This follows immediately from
to the GW relation, after the fermion integration is performed, 
as shown in the Appendix.
\item In the quenched limit the
r.h.s.\ of eq.~(\ref{eq:AWI_ZV}) possesses a double pole at
vanishing quark (pion) mass with a residue related to
$m^2_{\eta'}$~\cite{WIT,VEN}. To be precise its residue is in our
normalization $m_{\eta'}^2 m_\pi^4 F_\pi^2/2N_f$~\footnote{We are
using the definition $\<0|\partial_\mu\widehat{\cal A}^0_\mu|\eta'\>=
\sqrt{2N_f}F_{\eta'}m^2_{\eta'}$ with the identification of
$F_{\eta'}$ with $F_{\pi}$ in the quenched limit. We are also
assuming that the pion bound state exists even in the quenched
theory.}. This observation was the basis of the many 
quenched simulations carried out in the years in lattice
QCD aimed at extracting the mass of the pseudo-scalar flavour
singlet, starting from the seminal work of Hamber, Marinari,
Parisi and Rebbi (see refs.~\cite{PAR,KUR,KIL}). \item With the
idea of trying to set up a formula for the $\eta'$ mass which
would not depend on the details of the lattice definition of the
topological charge density operator, in ref.~\cite{YOS} for Wilson
and few years later for staggered fermions~\cite{Smit:1986fn}, the
equation \beq \frac{m^2_{\eta'}F_\pi^2}{2N_f}\Big{|}_{N_f/N_c=0}=
\lim_{m\rightarrow 0} \frac{(2 m)^2}{(2N_f)^2} \int
d^4x\,\<P^0(x)P^0(0)\>\Big{|}_{\rm{quenched}}^{\rm{ZV}}
\label{eq:PPZVQ} \eeq was argued to hold in the limit
$N_f/N_c\rightarrow 0$~\footnote{Actually the way the $\eta'$ mass
formula was written in ref.~\cite{YOS} is wrong by the finite
normalization factor $1-\partial\bar{M}^0(M)/\partial M|_{M=M_{cr}}$, 
see ref.~\cite{BMMRT} for notations.}. Eq.~(\ref{eq:PPZVQ}) can be 
regarded as an expression of the
residue of the $1/m^2$ double pole present in the $\int
d^4x\,\<P^0(x)P^0(0)\>$ correlator at $N_f/N_c=0$ (i.e.\ in the
absence of the fermion determinant). In this way no use of the
identity~(\ref{eq:AWI_ZV}) is actually made, though
eq.~(\ref{eq:PPZVQ}) is obviously consistent
with~(\ref{eq:AWI_ZV}) for GW fermions.
\end{itemize}
\item As in the formal continuum theory~\cite{LS}, the chiral condensate 
can be extracted from the small $m$ expansion of the (lattice) full 
topological susceptibility, as defined by eqs.~(\ref{eq:TCDO}) 
and~(\ref{eq:AWI_P0Q})  
\be
4N_f^2\chi_{tL}^{\rm{full}}= -4m\<S^0(0)\>\Big{|}_{m=0}+{\mbox{O}}(m^2)\, ,
\label{eq:GORR}
\ee
with no need of power divergent subtractions.
\end{enumerate}

\section{Conclusions}
\label{sec:CONCL}

In this paper we have shown that there are no $m^2/a^2$ power 
divergences in $\chi_{tL}^{\rm{full}}$, as defined in 
eq.~(\ref{eq:AWI_P0Q}). Thus, if GW fermions are used, the topological 
susceptibility of the full theory
\begin{eqnarray}
\chi_{tL}^{\rm{full}}= a^{-6} \int d^4x\,\<\frac{1}{2} \Tr [\gamma_5 D(x,x)]
\frac{1}{2} \Tr [\gamma_5 D(0,0)]\>
\label{eq:TOPFUL}
\end{eqnarray}
can be employed to extract the physical value of the chiral condensate using 
the eq.~(\ref{eq:GORR}), without the need of performing any power subtraction. 
The reason is that the $a^{-2}$ power divergence in $\<S^0\>$, 
outside the chiral limit, is exactly
compensated by a similar divergence in the $\<P^0P^0\>$ 
correlator in the r.h.s.\ of eq.~(\ref{eq:AWI_P0Q}).

Actually we have proved more than that. We have proved that all
$a^{-2}$ power divergences in the $\<P^0P^0\>$ correlator arise
from the ZC contributions. This conclusion follows from the
non-singlet WTI~(\ref{eq:AWI_Pf}), where we see that only ZC
diagrams contribute.

This observation may be of some interest in view of the
formula~(\ref{eq:PPZVQ}), where the (quenched) $\eta'$ mass is
expressed in terms of the ZV quenched correlator of two singlet
pseudo-scalar quark densities. In this case, in order to conclude that power
divergences are also absent from the r.h.s.\ of eq.~(\ref{eq:PPZVQ}),
one must assume that taking the limit $N_f/N_c\rightarrow 0$ in
eq.~(\ref{eq:AWI_ZV}) is equivalent to dropping the fermion
determinant  and does not introduce unexpected $1/a^2$-power 
divergences. If this is the case, one can imagine to check
the validity of the formula~(\ref{eq:WVFOR}) and the assumptions
underlying it by comparing the value of the YM topological
susceptibility (r.h.s.\ of eq.~(\ref{eq:ETAMASSIND})) with the
residue of the double pole that arises at zero quark mass when the
fermion determinant is dropped from eq.~(\ref{eq:AWI_ZV}) (thus
ending up with eq.~(\ref{eq:PPZVQ})). In this context it is 
interesting to mention that the YM topological 
susceptibility has been recently computed at several 
values of the lattice spacing by counting the number of zero modes 
of the Neuberger-Dirac operator~\cite{Giusti:2003gf,DelDebbio:2003rn}.
Data are compatible with the scaling behaviour expected for
a quantity of dimension $d=4$ and 
no sign of power divergences (within errors).
\section*{\bf Appendix} 
\label{sec:APP}

In this appendix we want to prove that the formula 
\begin{eqnarray}
\frac{N_f}{2a^3}\int d^4x\,\<\< \Tr [\gamma_5 D(x,x)]\>\>=
m\int d^4x\,\<\<P^0 (x,x)\>\>\, ,
\label{eq:INTER}
\end{eqnarray}
where $\<\<\ldots\>\>$ means fermion field contraction only, holds 
configuration by configuration in the massive theory.
To this end we recall that the expression of the massive GW operator 
in eq.~(\ref{eq:action}) is
\begin{eqnarray}
D_m=(1-\frac{am}{2})D+m\, .
\label{eq:MASSDGW}
\end{eqnarray}
In terms of $D_m$
the GW relation~(\ref{eq:GW}) takes the form
\begin{eqnarray}
\gamma_5D_m+D_m\gamma_5-2m\gamma_5=\displaystyle{\frac{a}{(1-\frac{am}{2})}}
\Big{(}D_m\gamma_5D_m-m(\gamma_5D_m+D_m\gamma_5)+m^2\gamma_5\Big{)}\, .
\label{eq:MASSRGW}
\end{eqnarray}
Putting similar terms together and multiplying by the inverse of 
the massive GW operatore, $D_m^{-1}$, we get from~(\ref{eq:MASSRGW}), 
after taking the trace and integrating over space-time
\begin{eqnarray}
-m\int d^4x\Tr[\gamma_5D_m^{-1}]=\frac{a}{2}\int d^4x\Tr[\gamma_5D_m]\, .
\label{eq:INTERM}
\end{eqnarray}
At this point it is enough to observe that
 
1) contracting the fermion 
fields in the operator $P^0$ appearing in eq.~(\ref{eq:INTER}) gives
\begin{eqnarray}
-\frac{1}{N_f}\int d^4x\,\<\<P^0 (x,x)\>\>=\frac{1}{a^4(1-\frac{am}{2})}
\int d^4x\Tr[\gamma_5D_m^{-1}]\, ,
\label{eq:PD}
\end{eqnarray}

2) tracing $D_m$ with $\gamma_5$ 
yields 
\begin{eqnarray}
\Tr[\gamma_5 D_m]=(1-\frac{am}{2})\Tr[\gamma_5 D]\, .
\label{eq:QDGW}
\end{eqnarray}

\vskip .3cm
\section*{\bf Acknowledgments} We wish to thank M. L\"uscher and
G. Veneziano for useful discussions. This work was supported in part by the
European Community's Human Potential Programme under contract
HPRN-CT-2000-00145, Hadrons/Lattice QCD.

\end{document}